\documentclass[journal=ancac3,manuscript=article]{achemso}
\usepackage{hyperref}
\usepackage[USenglish]{babel}
\usepackage{hyphenat}
\usepackage{xspace}
\usepackage[dvipsnames]{xcolor}
\usepackage[version=4]{mhchem}
\usepackage{caption}
\usepackage{subcaption}
\usepackage{siunitx}[=v2]
\newcommand{\qty}[2]{\SI{#1}{#2}}
\newcommand{\qtyrange}[3]{\SIrange{#1}{#2}{#3}}

\newcommand{\unit}[1]{\si{#1}}
\DeclareSIUnit\Ang{\angstrom}
\DeclareSIUnit\Ry{Ry}
\DeclareSIUnit\layer{layer}
\DeclareSIUnit\ac{\textit{e}}
\DeclareSIUnit\atom{atom}
\DeclareSIUnit\Kps{\kelvin\per\pico\second}
\DeclareSIUnit\dens{\gram\per\centi\metre\cubed}
\DeclareSIUnit\surf{\eV\per\Ang\squared}
\usepackage[capitalise,noabbrev]{cleveref}
\newcommand{\SuppInf}{Supporting Information\xspace}

\newcommand{\TiSi}{\ce{TiSi_2}\xspace}
\newcommand{\TiSix}{\ensuremath{\ce{TiSi_{\mathit{x}}}}\xspace}

\newcommand{\cSi}{\ensuremath{\alpha\text{-Si}}\xspace}
\newcommand{\CIVIX}{\ensuremath{\text{C49-}\ce{TiSi2}}\xspace}
\newcommand{\CVIV}{\ensuremath{\text{C54-}\ce{TiSi2}}\xspace}
\newcommand{\AAi}{\ensuremath{\text{AA}_\mathrm{i}}\xspace}
\newcommand{\CCi}{\ensuremath{\text{CC}_\mathrm{i}}\xspace}

\newcommand{\mySurf}{\ensuremath{c(4\times2)~\ce{Si}(100)}\xspace}

\newcommand{\Eform}{\ensuremath{E_\text{form}}\xspace}
\newcommand{\Eads}{\ensuremath{E_\text{ads}}\xspace}
\newcommand{\Eassc}{\ensuremath{E_\text{assc}}\xspace}
\newcommand{\Esurf}{\ensuremath{E_\text{surf}}\xspace}

\renewcommand{\i}{\ensuremath{\ce{Ti}^{(\mathbf{1})}}\xspace}

\newcommand{\iii}{\ensuremath{\ce{Ti}^{(\mathbf{3})}}\xspace}
\newcommand{\iv}{\ensuremath{\ce{Ti}^{(\mathbf{4})}}\xspace}
\newcommand{\Tiv}{\ensuremath{\ce{Ti}^{(\mathbf{5})}}\xspace}
\newcommand{\vi}{\ensuremath{\ce{Ti}^{(\mathbf{6})}}\xspace}
\title{%
Topology-Directed Silicide Formation: \\
An Explanation for the Growth \\
of C49-\ce{TiSi2} on the Si(100) Surface}
\author{Lukas H{\"u}ckmann}
\affiliation{Leiden Institute of Chemistry, Gorlaeus Laboratories, Leiden University, P.O. Box 9502, 2300 RA Leiden, The Netherlands}
\alsoaffiliation{Advanced Research Center for Nanolithography, Science Park 106, 1098 XG Amsterdam, The Netherlands}
\alsoaffiliation{Institute for Theoretical Physics, University of Amsterdam, Postbus 94485, 1090 GL Amsterdam, The Netherlands}
\author{Jonathon Cottom}
\affiliation{Advanced Research Center for Nanolithography, Science Park 106, 1098 XG Amsterdam, The Netherlands}
\alsoaffiliation{Institute for Theoretical Physics, University of Amsterdam, Postbus 94485, 1090 GL Amsterdam, The Netherlands}
\alsoaffiliation{Leiden Institute of Chemistry, Gorlaeus Laboratories, Leiden University, P.O. Box 9502, 2300 RA Leiden, The Netherlands}
\author{J{\"o}rg Meyer}
\affiliation{Leiden Institute of Chemistry, Gorlaeus Laboratories, Leiden University, P.O. Box 9502, 2300 RA Leiden, The Netherlands}
\author{Emilia Olsson}
\email{k.i.e.olsson@uva.nl}
\alsoaffiliation{Advanced Research Center for Nanolithography, Science Park 106, 1098 XG Amsterdam, The Netherlands}
\affiliation{Institute for Theoretical Physics, University of Amsterdam, Postbus 94485, 1090 GL Amsterdam, The Netherlands}
\date{\today}

\begin{document}

\begin{abstract}
Designing metal-semiconductor junctions is essential for optimizing the performance of modern nanoelectronic devices. 
A widely used material is \TiSi, which combines low electronic resistivity with good endurance. 
However, its multitude of polymorphs continues to pose a challenge for device fabrication. 
In particular, the naturally occurring formation of the metastable \CIVIX modification remains poorly understood and is problematic due to its unfavorable electronic properties.
Based on extensive DFT calculations, we present a comprehensive model of \ce{Ti} adsorption on \ce{Si(100)} that highlights the pivotal role of surface topology for the initial stages of the interfacial \ce{TiSi2} formation process.
We show that the interplay between \ce{Si} surface dimers, the symmetry of the \ce{Si(100)} surface, and the incorporation of \ce{Ti} adsorbates below the surface drives an adsorption pattern that yields a nucleation template for the \CIVIX phase. 
Our atomistic model rationalizes experimental observations like the Stranski-Krastanov growth mode, the preferential formation of \CIVIX despite it being less favorable than the competing C54 phase, and why disruption of the surface structure restores thermodynamically driven growth of the latter. 
Ultimately, this novel perspective on the unique growth of \TiSi will help to pave the way for next-generation electronic devices.
\end{abstract}

\section{Introduction}
\label{sec:intro}

The performance of modern micro- and optoelectronic devices hinges on reliable metal–semi\-conductor contacts at nanometer scales. Among available contact materials, titanium disilicide (\TiSi) is widely used in Si-based architectures due to its low electrical resistivity, small Schottky barrier on n-type Si, and chemical stability.\cite{sze2021,Zhang2003,Zhang2012, zhang2020} These properties have established \TiSi\ as an important material for CMOS technologies, Schottky diodes, and photodetectors,\cite{de_bosscher1986,dheurle1998,pelleg2004,wang2015,asil_ugurlu2021} and have prompted exploration of its role in emerging applications, including photocatalytic water splitting\cite{ritterskamp2007, lin2009,peng2024, hannula_highly_2019} and perovskite-\ce{Si} tandem photovoltaics.\cite{pyun2024,yu2009}

Yet, despite its technological ubiquity, the fabrication of phase-pure \TiSi remains a persistent challenge. Upon annealing, Ti deposited on crystalline Si forms a metastable orthorhombic C49 phase, which only transforms to the desired low-resistivity C54 polymorph at elevated temperatures.\cite{zhang2004,beyers1985,kematick1996,ting1986} The persistent nucleation of C49-\TiSi — even under conditions where C54 is thermodynamically favored\cite{kasica1997} — suggests a complex and poorly understood interplay of interfacial energetics, surface mobility, and local strain. The limitations of C49 not only degrades the electrical performance through increased contact resistance, but also impose stringent thermal budgets that constrain integration with temperature-sensitive materials in advanced device stacks,\cite{ting1986,lu1991,asil_ugurlu2021} such as 3D architectures and backside power delivery networks.\cite{Xie_2024,Jaydeep_23,Kobrinsky_2023} Understanding and ultimately controlling this phase selection is thus critical for enabling next-generation nanoscale logic\cite{gao_synaptic_2023, deo_super-turing_2025, lee_hfzro-based_2025,nathan_si-based_2022} and optoelectronic platforms.\cite{yadav_substrate_2019,pyun2024,peng2024, cheng_synergistic_2025,hannula_fabrication_2016}

In standard device processing, thin \TiSi layers are formed via self-aligned silicidation (salicidation), where Ti is deposited onto single-crystalline Si and subsequently annealed.\cite{mallardeau1989} Initial Ti growth follows a Stranski-Krastanov-like mode:\cite{stranski1937,baskaran2012} after formation of a monolayer, additional adatoms cluster into three-dimensional islands. Notably, XPS measurements reveal the onset of silicide formation even at room temperature,\cite{vahakangas1986,palacio2007} suggesting immediate interfacial reactivity upon deposition. Upon thermal activation, Ti diffuses into the Si substrate and reacts to form silicide phases. The binary Si–Ti system exhibits a rich phase diagram with multiple low-symmetry compounds and stoichiometries,\cite{yamane1994,bulanova2003} with the resulting structure being dictated by both temperature and annealing duration. At approximately \qty{600}{\degreeCelsius}, the orthorhombic C49-\ce{TiSi2} polymorph (space group $Cmcm$) nucleates, which transforms into the stable C54 phase ($Fddd$) above \qty{700}{\degreeCelsius}.\cite{mann1994} The C54 polymorph is technologically important due to its lower resistivity (15–20~\si{\micro\ohm\cm} vs. 60–80~\si{\micro\ohm\cm} for C49)\cite{beyers1985} and higher thermodynamic stability compared to C49 ($\Delta\Delta H_f \approx 2$–$4~\si{\kilo\joule\per\mol}$).\cite{kematick1996,kasica1997}

Despite these insights, salicidation invariably produces \CIVIX as the initial phase,\cite{zhang2004} necessitating complex, energy-intensive multi-step annealing procedures to induce C54 formation and heal microstructural defects that would otherwise compromise yield and long-term reliability. The unresolved origin of the detrimental formation sequence of these two phases represents a significant bottleneck in fully harnessing \TiSi's potential for scaled technologies. Various studies have attributed the preference for C49 to its lower surface energy\cite{jeon1992,wang2006} or enhanced Ti diffusivity within its lattice,\cite{motakef1991} indicating a kinetically driven process. Yet, these hypotheses remain inconclusive, and a consistent atomic-scale explanation for first forming the C49 phase is lacking.\cite{Zhang2003,brown2021}

To address this gap, several atomistic simulation studies have examined \TiSi formation at the Si interface. Early work by \citet{yu1998b} modeled Ti deposition on an unreconstructed $1\times1$ Si(100) surface and reported that interfacial energetics favor C49-\ce{TiSi2} over C54. Subsequent studies investigated Ti adsorption and adatom diffusion,\cite{yu1998,briquet_first_2013,anez2012} showing that the Si(100) surface reconstruction with the surface atoms forming dimers strongly influences adatom clustering and adsorption energetics, though high surface coverages and silicide intermixing were not investigated. More recent efforts have focused on the bulk and defect thermodynamics: \citet{sanchez2009} assessed the formation energies of Ti defects in crystalline Si, while \citet{brown2021} systematically evaluated the phase stability of the C49 and C54 polymorphs. 
Collectively, these studies underscore the delicate energetic balance at the Si–Ti interface, but an understanding of how atomic-scale processes drive C49 nucleation remains elusive. This phenomenon exemplifies a broader class of interface-driven metastability in thin-film heterostructures, where transient polymorphs emerge and persist despite being thermodynamically disfavored -- a recurring theme across material systems such as silicides, nitrides, and transition-metal chalcogenides.\cite{madan1997_cAlN,kim2001_cAlN,tang2018_mote2_apl,tao2022_mote2_perspective,prabhu2020}

In this paper, we perform density functional theory calculations to develop a comprehensive model for the initial stages of Ti growth on Si(100). Starting from isolated Ti adsorbates and extending to coverages of two monolayers, we systematically identify preferred adsorption sites, interaction mechanisms, and the conditions for which adsorbed \ce{Ti-Ti} pairs lift the \mySurf surface reconstruction. We show that these local structural motifs disrupt Si dimers and drive the self-promoting formation of a TiSi bilayer. Remarkably, this bilayer reproduces the low-symmetry zigzag motifs unique to the C49 polymorph, thereby acting as a nucleation template for its growth. 
Our findings reveal that the emergence of C49 is governed primarily by surface topology rather than by enhanced diffusion or interfacial stress as previously assumed.
This provides a unified framework for understanding Stranski-Krastanov growth and paves the way towards suppression the C49 phase by surface disruption, preamorphization, or refractory metal overlayers.

\section{Results and Discussion}
\label{sec:Res}
\subsection{Ti-Deposition}
\label{sec:single}
\begin{figure}
    \centering
    \includegraphics[width=\textwidth]{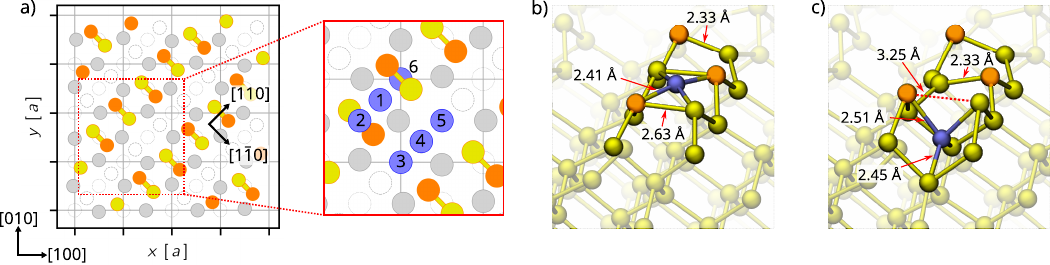}
    \caption{a) Schematic of the \ce{Ti} adsorption sites on the \mySurf surface (top view). The numbered blue circles in the inset represent Ti adatoms, the orange and yellow ones represent buckled Si-dimers with the orange ones being elevated, and the gray, the dashed-hollow, and the dotted hollow are the Si atoms of first, second, and third subsurface layer, respectively.  Depiction of a Ti atom adsorbed at position b) \textbf{1} and c) \textbf{6}, respectively, where Si is colored yellow, Ti  blue.}
    \label{fig:single_Ti}
\end{figure}

The pristine Si(100) surface consists of two surface silicon atoms per surface unit cell, each with two-fold coordination. To maximize coordination%
, these surface atoms reconstruct by tilting and forming dimers with neighboring atoms. The resulting Si dimers are not perfectly parallel to the surface but exhibit an asymmetric buckling, giving rise to various possible surface symmetries. 
Comparing the energetics of different surface configurations confirms that the $p(2\times2)$ and $c(4\times2)$ reconstructions represent the lowest-energy configurations, consistent with STM measurements~\cite{wolkow1992,hata2002} and earlier computational studies.\cite{ramstad1995,guo2014}
The Si dimers align parallel to each other along the [110]-direction on the surface forming characteristic patterns of Si-dimer ridges and trenches (see \cref{fig:single_Ti}a, yellow/orange sites vs. hollow circles). Further details on the surface reconstructions are provided in the \SuppInf, Section~1.

\begin{table}
    \centering
    \caption{Adsorption energies (\Eads) of isolated Ti atoms on Si(100) of the different sites on the \mySurf surface.}
    \label{tab:Ti-Eads}
    \begin{tabular}{lr}
    \hline
    site & \Eads $[\,\unit{\eV\per\atom}\,]$ \\
    \hline
    \textbf{1} & $-\num{0.06}$ \\
    \textbf{2} & $\num{2.19}$ \\
    \textbf{3} & $\num{0.65}$ \\
    \textbf{4} & $\mathbf{4}\to\mathbf{3}$ \\
    \textbf{5} & $\num{0.20}$ \\
    \textbf{6} & $-\num{0.14}$ \\
    \hline
    \end{tabular}
\end{table}

On the \mySurf surface, five high-symmetry adsorption sites can be identified, as illustrated in \cref{fig:single_Ti}a: The four-fold coordinated hollow site between two dimers (\textbf{1}), on a bridge site centered on a buckled dimer (\textbf{2}), and three trench sites between dimers (\textbf{3}–\textbf{5}).  Among these, only \i is energetically favorable, with $\Eads(\i) = \qty{-0.06}{\eV}$,  while all the others are significantly higher in energy, whereby \iv is unstable and relaxes to \iii during geometry optimization (see \cref{tab:Ti-Eads}). 

At site \i, the Si dimers align and slightly open, increasing the \ce{Si-Si} bond length from \qty{2.33}{\Ang} to \qty{2.63}{\Ang}. The Ti atom symmetrically coordinates to neighboring Si atoms, forming four identical bonds of \qty{2.41}{\Ang} (see \cref{fig:single_Ti}b). In contrast, sites \iii and \Tiv favor tetrahedral coordination, with the ideal geometry hindered by the rigidity of the surrounding surface atoms. Embedding Ti into a perfect tetrahedral arrangement would necessitate breaking or significantly distorting adjacent Si dimers. Such rearrangement is energetically unfavorable, as it would leave opposing Si atoms undercoordinated. As a result, the \ce{TiSi4} tetrahedron remains distorted at trench sites, with varying degrees of distortion at \iii and \Tiv giving rise to their unfavorable adsorption energies \Eads (\qty{0.65}{\eV} and \qty{0.20}{\eV}). This structural constraint explains why trench sites are less favorable than the top site. Indeed, as will be demonstrated in the following sections, the Ti adsorption energetics are intrinsically linked to the opening of Si dimers.

Furthermore, without considering any potential barriers for reaching this site, Ti can also submerge beneath a dimer (site \textbf{6}, \cref{fig:single_Ti}c), adopting a relatively unperturbed tetrahedral environment. The \ce{Ti-Si} bond lengths range from \qtyrange{2.45}{2.51}{\Ang}, with bond angles spanning from \ang{81} to \ang{121}. The overlying Si dimer is stretched ($r_{\ce{SiSi}} = \qty{3.25}{\Ang}$) and slightly elevated above the surface ($\Delta z = \qty{0.13}{\Ang}$). However, no extrusion of surface atoms occurs in this scenario, in contrast with findings by \citet{yu1998} and observations made in the context of silicidation of other transition metals.\cite{cottom2024} From a thermodynamic point of view, further embedding of Ti into the Si slab does not occur, as the formation energy of Ti interstitials in bulk Si is highly unfavorable compared to subsurface adsorption layers (see \SuppInf, Section~2). These observations align closely with previous studies~\cite{yu1998, anez2012, briquet_first_2013} in terms of geometric relaxations and adsorption energetics. Additionally, these earlier works demonstrated that Ti readily migrates across the surface. Based on these findings, the \i site will be preferentially occupied at low coverages, providing the starting point for subsequent growth of the first monolayer.

\subsection{Interaction between adsorbed Ti pairs}
\label{sec:pairs}

\begin{figure}
    \centering
    \includegraphics[width=.5\linewidth]{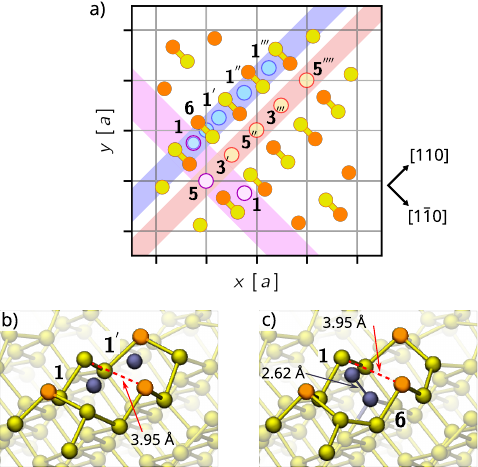}
    \caption{a) Sketch of the \mySurf surface to illustrate the direction-dependent pairwise arrangement of Ti adsorbates with the sites on the Si dimer ridge being colored blue, those in the trench being red, and those across along the [1$\bar{1}$0]-direction being purple. The colored stripes are added to guide the eye. Subsurface Si atoms are omitted for visibility. %
    b)/c) Depiction of two Ti atoms at position b) $\i$-$\ce{Ti}^{(\mathbf{1}')}$ and c) $\i$-$\vi$, respectively, where Si is colored yellow, and Ti is blue.}
    \label{fig:Tipairs}
\end{figure}

\begin{table}
    \centering
    \caption{The adsorption and association energies (\Eads, \Eassc) of $2\times\ce{Ti}@\ce{Si(100)}$. The sampling of the Ti pairs has been done in accordance with the symmetry of the Si surface leading to three sets of pairs: On the Si dimer ridge in [110] direction ($\mathbf{1}$-$\mathbf{1}^x$), along the trench in [110]-direction ($\mathbf{5}$-$\mathbf{3}/\mathbf{5}^x$), and across along the [1$\bar{1}$0]-direction ($\mathbf{1}$-$\mathbf{1}/\mathbf{5}^x$). Compare also \cref{fig:Tipairs}a. First neighbors to the reference atom, second neighbors, and so on  are denoted as ${'}$,${''}$,...~.}
    \label{tab:Ti2}
    \begin{tabular}{lrr| lrr| lrr}
    \hline
    $\mathbf{1}$ & $\Eads$ & \Eassc  & $\mathbf{5}$ & $\Eads$  & \Eassc  & $\mathbf{1}$ & $\Eads$  & \Eassc \\
     $[110]$ & [\unit{\eV\per\atom}] & [\unit{\eV}] & [110] & [\unit{\eV\per\atom}] & [\unit{\eV}] & [1$\bar{1}$0] & [\unit{\eV\per\atom}] & [\unit{\eV}] \\
    \hline
    $\mathbf{1}^{'}$ & \num{-0.04} & \num{0.04} & $\mathbf{3}^{'}$ & \num{0.20} & \num{-0.45} & $\mathbf{5}^{'}$ & \num{0.12} & \num{0.10} \\
    $\mathbf{1}^{''}$ & \num{-0.05} & \num{0.02} & $\mathbf{5}^{''}$ & \num{0.28} & \num{0.15} & $\mathbf{1}^{''}$ & \num{-0.05} & \num{0.02}\\
    $\mathbf{1}^{'''}$ & \num{-0.01} & \num{0.11} & $\mathbf{3}^{'''}$ & \num{0.55} & \num{0.24} & \\
    $\mathbf{6}$ & \num{-0.18} & \num{-0.16} & $\mathbf{5}^{''''}$ & \num{0.26} & \num{0.11} & \\
    \hline
    \end{tabular}
\end{table}

To confirm that synergistic effects do not alter the preference for the \i site, we now first investigate potential stabilization of the other sites via adsorbate pair formation.
To systematically explore these interactions, they are categorized based on their orientation: along the [110]-direction on the Si dimer ridges, along the [110]-direction parallel to the trenches between the Si dimer ridges, and along the [1$\bar{1}$0]-direction across the ridges and trenches (see \cref{fig:Tipairs}a).

When a second Ti atom ($\ce{Ti}^{(\mathbf{1}')}$) adsorbs along the same dimer ridge at closed possible distance, both Ti atoms bind simultaneously to the Si atoms of the shared surface dimer in between. As a result, the dimer reconstruction locally lifts, fully opening the Si–Si bond to \qty{3.95}{\Ang}, matching the spacing found on the unreconstructed Si(100) surface (see \cref{fig:Tipairs}c). This lateral interaction slightly disfavors co-adsorption as indicated by an association energy $\Eassc = \qty{+0.04}{\eV}$ (\cref{tab:Ti2}). In contrast, for the second and third nearest neighbor configurations ($\ce{Ti}^{(\mathbf{1}'')}, \ce{Ti}^{(\mathbf{1}''')}$), each Ti atom independently relaxes the local environment, nearly identical to isolated defects. However, minor residual interactions remain, reflected in small but positive association energies of \qty{0.02}{\eV} and \qty{0.11}{\eV}, respectively.

Similarly, Ti adsorbates at trench sites exhibit the same general trend: their pairwise association energies (\Eassc) are positive, meaning additional Ti occupation in the vicinity of the first adsorbate further reduces the accessibility of these already unfavorable positions. An exception occurs for the directly neighboring \iii-\Tiv pairs, for which \Eassc is \qty{-0.45}{\eV}. However, even this negative \Eassc is insufficient to offset the intrinsically high \Eads values of these isolated defects (\cref{tab:Ti2}), resulting in a net adsorption energy of \qty{0.20}{\eV\per\atom}. Likewise, an adjacent top-site adsorbate at position \i does not provide stabilization when paired with site \Tiv, as indicated by an \Eassc of \qty{0.10}{\eV}. In this configuration, the presence of Ti at \i restricts the flexibility of the adjacent Si dimer, hindering the formation of an optimal coordination environment at \Tiv. Consequently, clustering of Ti adsorbates at any symmetry site is energetically disfavored due to mutual interference as each atom seeks a high coordination number. The associated energetic penalty for neighboring Ti atoms aligns with values reported previously,\cite{briquet_first_2013} providing a driving force toward the uniform initial growth of the titanium layer.

However, there is one notable exception to the trend discussed above: the interaction between sites \i and \vi. This pair is highly favorable, with $\Eads=\qty{-0.18}{\eV\per\atom}$ and $\Eassc=\qty{-0.16}{\eV}$, due to the formation of a short \ce{Ti-Ti} bond measuring \qty{2.62}{\Ang} (see \cref{fig:Tipairs}c). While the coordination of \vi remains tetrahedral, similar to the isolated defect, the \i site abandons its previously described $D_{4h}$-like coordination and moves closer to the neighboring Ti atom. The Si–Si dimer above \vi remains fully opened, with Si atoms effectively returning to the unreconstructed configuration. This result implies that there exists a stable and energetically favorable local configuration where the pristine Si(100) surface reconstruction is locally reversed. As discussed in detail in the following section, the surface progressively transforms when the number of \i-\vi pairs increases, eliminating the initial distinction between trench and dimer-ridge sites. 

\subsection{Increasing Coverages}

\begin{figure}[htb!]
    \centering
    \includegraphics[width=\textwidth]{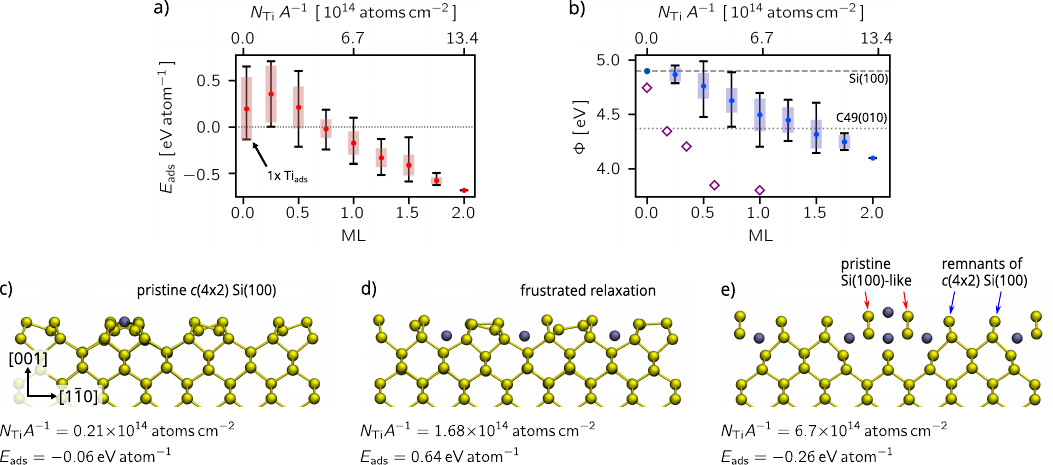}
    \caption{a) Adsorption energy (\Eads) and b) the work function $\Phi$ as a function on the  Ti coverage. The adsorbate coverage is given in monolayers (ML) defined via the number of adsorbates per surface unit cell and in atoms per surface area ($N_\mathrm{Ti}A^{-1})$. The error bars indicate the highest/lowest \Eads and the boxes are the standard deviation of \Eads of the ensemble of configurations. 
    For comparison, the $\Phi$ of pristine Si(100) and \CIVIX(010) is added as dashed and dotted line, respectively. The hollow diamonds are UPS data measured by Taubenblatt and Helms.\cite{taubenblatt_silicide_1982} c)-e) Three representative examples of the slab in c) the dilute limit, d) at low concentration, and e) at medium coverage (from left to right). All slabs are displayed along the [110]-axis. }
    \label{fig:NxTi}
\end{figure}

Higher coverages and the resulting combinatorial sampling problem has been approached by accounting all symmetrically unique Ti$_\mathrm{ads}$ configurations within the third neighbor sphere in accordance with the previously characterized lateral interactions. Expanding these finite-sized configurations to the $4\times4$ slab results in step-wise coverage increases of 0.25\,MLs ($\hat{=}\qty{8}{atoms}$), see Methods for more details.

When increasing the number of Ti adsorbates, a pronounced non-monotonic dependence of the average adsorption energy per adsorbate on coverage emerges, see \cref{fig:NxTi}a. At low coverages, most configurations yield an unfavorable \Eads, peaking at 0.25 monolayers (MLs) with on average \qty{0.35}{\eV\per\atom} and a maximum outlier reaching up to \qty{0.71}{\eV\per\atom}. This behavior can be understood by the interplay of isolated and paired adsorbates described above: at low coverages, 
\Eads is strongly influenced by the number of Ti atoms occupying trench sites and the associated pair-wise interactions, and whether the Si dimers are opened in a stable or unstable fashion. Low adsorbate concentrations allow for many unfavorable configurations simply by combinatorics, and the disadvantageous effects of Ti pairs identified in the previous section can amplify one another. An example for such a configuration is displayed in \cref{fig:NxTi}d, where eight \iii and \Tiv atoms collectively attempt to pull neighboring buckled Si dimers to achieve higher coordination. However, the lack of adatoms between the dimers prevents this relaxation, resulting in an overall frustrated geometry (compare also the pristine dimers in \cref{fig:NxTi}c).   
The observed spread in \Eads across all coverages is therefore a result of the complex interplay of opened Si dimers with or without stabilization by neighboring Ti atoms, nearest-neighbor effects as discussed in the previous section, and the occupation of trench sites prior to the surface reconstruction being lifted. While these contributions cannot be easily separated, highly ordered configurations support this interpretation, e.g., exclusive occupation of trench sites yields highly unfavorable \Eads (\qty{0.25}{ML}, $\Eads=\qty{0.71}{\eV\per\atom}$), while perfectly evenly distributed \i adsorbates with maximized pair distances yields moderate values (\qty{0.25}{ML}, $\Eads=\qty{-0.08}{\eV\per\atom}$). 

At coverages above 0.75\,ML, \Eads rapidly becomes negative due to the increasing number of \i-\vi pairs, which yield more favorable nearest-neighbor interactions (compare \cref{fig:Tipairs}c). More importantly, these pairs locally lift the \mySurf reconstruction and stabilize the Si atoms in a configuration that is identical to the unrelaxed Si(100) surface, thereby eliminating the distinction between Si dimer ridge and trench sites. As a result, the previously unfavorable \iii,\Tiv sites become equivalent to the more favorable \i,\vi sites. This is exemplified in \cref{fig:NxTi}e: A stack of \i-\vi pairs in the center opens the surface dimers and pushes the adjacent Si atoms back into their original crystallographic positions. This, in turn, stabilizes neighboring Ti adsorbates and enables further Ti deposition on the 4-fold hollow sites between the leveled dimers. Since this effect is local, remnants of the \mySurf reconstruction are still discernible in the second-nearest row.
Eventually, at coverages close to 2\,ML, the surface reconstruction is fully reversed and consists only of  \i, \vi configurations. This transformation reduces the adsorption energies to \qty{-0.69}{\eV\per\atom} at full occupation of all \i, \iv sites.  

This has several important consequences. First, successive occupation of \i-\vi pairs implies that surface and subsurface sites fill concurrently, even in the dilute limit. A TiSi bilayer therefore nucleates concomitantly rather than only after completing a well-defined Ti monolayer. The calculated work function (\(\Phi\)) decreases smoothly from \qty{4.87}{\eV} for pristine Si(100) to \qty{4.10}{\eV} for the completed TiSi bilayer. These values compare closely with UPS measurements by Taubenblatt and Helms (\qty{4.76}{\eV} and \qty{3.80}{\eV}), with absolute deviations of \qty{0.11}{\eV} (bare) and \qty{0.30}{\eV} (bilayer), i.e. \(\approx\)2.3\,\% to 7.9\,\% relative to experiment.\cite{taubenblatt_silicide_1982} UPS probes macroscopic samples with steps, partial reconstructions, varying termination, thermal disorder, and spatially varying Ti coverages. Our calculations on the other hand are performed on a perfectly ordered slab, which rationalizes these small relative differences.

More importantly, the formation of the TiSi bilayer is a self-promoting process. The gradual opening of Si dimers and pairing \i-\vi sites also gradually lifts the surface reconstruction, eliminating the distinction between trench and ridge site and leaving only the energetically favorable \i,\vi sites. Consequently, each \i-\vi pair creates additional \i,\vi sites, driving a positive feedback loop where Ti adsorption becomes increasingly favorable with coverage. Thus, the deposition of Ti and the initial formation of a mixed \ce{TiSi} layer is not solely explained by adsorption energies of individual atoms alone, but by the complex interplay of Ti-Ti pairing and opening of the Si surface dimers. Ultimately, the structure of the TiSi bilayer is dictated by how the \i–\vi\ pairs are embedded into the Si(100) surface.

\subsection{Characterization of the TiSi bilayer}

\begin{figure}
    \centering
    \includegraphics[width=\textwidth]{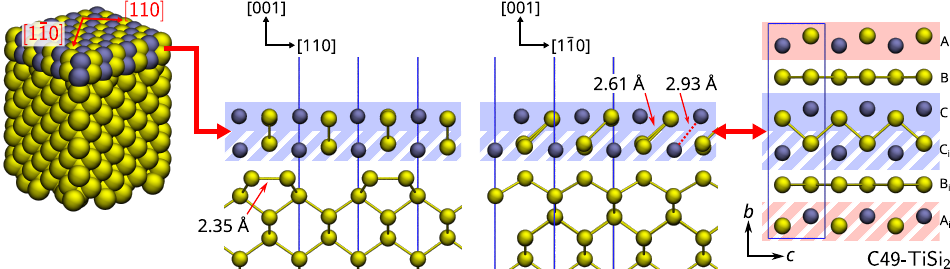}
    \caption{Depiction of a slab covered by two monolayers of Ti, with close-up views of the bilayer along the $[110]$ and $[1\bar{1}0]$ directions, together with a $3\times1\times3$ \CIVIX cell for comparison. The blue lines outline the $1\times1$ Si(100) surface unit cell.}
    \label{fig:C49}
\end{figure}

The fully occupied TiSi bilayer consists of alternating zigzag-shaped \ce{Ti-Ti} and \ce{Si-Si} chains oriented along the [110] lateral direction in the Si(001) surface (see \cref{fig:C49}). This arrangement yields local coordination environments that differ markedly from the $T_d$ motif in \cSi and the $D_{3h}$ motif in $\alpha$-Ti. The Si--Si bonds are elongated to $r_{\ce{SiSi}}=\qty{2.61\pm0.12}{\Ang}$ relative to their equilibrium length of \qty{2.35}{\Ang} in \cSi,\cite{Hubbard1975} while the Si--Si--Si angles are compressed to $\theta_{\ce{Si}}=\qty{95.28\pm2.75}{\degree}$. Minor out-of-plane corrugations in the bilayer lead to small variations in bond lengths and angles (see \cref{tab:bilayer}). It is also noteworthy that the underlying Si interface layer reconstructs to form Si--Si dimers similar to those on the Si(100) with the dimer buckling been lifted due to stabilizing interactions with the TiSi bilayer above.

\begin{table}
    \centering
    \begin{tabular}{llrrr}
    \hline
         && TiSi bilayer & \CIVIX & \cSi/$\alpha$-Ti \\
    \hline
    $r_{\ce{Ti-Ti}}$  & \unit{\Ang} & \num{2.93\pm0.15} & \num{3.32} & \num{2.92} \\
    $r_{\ce{Si-Si}}$  & \unit{\Ang} & \num{2.61\pm0.12} & \num{2.40} & \num{2.36} \\
    $\theta_{\ce{Ti}}$ & \unit{\deg} & \num{82.29\pm1.03} & \num{65.02} & \num{60} \\
    $\theta_{\ce{Si}}$ & \unit{\deg} & \num{95.28\pm2.75} & \num{96.27} & \num{109.47} \\
    $q_\mathrm{bader}(\ce{Ti})$ & \unit{\ac} & \num{0.96} & \num{1.07} & -- \\
    $q_\mathrm{bader}(\ce{Si_{A/C}})$ & \unit{\ac} & \num{-0.82} & \num{-0.53} & -- \\
    \hline
    \end{tabular}
    \caption{Bond lengths ($r$), angles ($\theta$), and Bader charges ($q_\mathrm{bader}$) in the TiSi bilayer on the Si(100) surface, in the \AAi\ layer of bulk \CIVIX, and in crystalline \cSi and $\alpha$-Ti, respectively. All data are from DFT calculations at the PBE level.}
    \label{tab:bilayer}
\end{table}

The comparison to \TiSix\ polymorphs from the literature with the TiSi bilayer reveals a remarkable structural similarity to the \CIVIX\ phase, particularly to the \AAi\ and \CCi\ layers (see \cref{fig:C49}). Note that the \AAi\ layer shown in \cref{fig:C49} is symmetrically equivalent to the \CCi\ layer; a more detailed description of the \CIVIX\ phase is provided in the \SuppInf, Section~1.3. Most strikingly, both the TiSi bilayer and the C49 structure feature zigzag chains of Ti and Si atoms. Such a motif is absent in other \TiSix polymorphs, especially in the competing C54 phase, where the Ti atoms are spaced out more evenly and thus do not align with the Ti-rich TiSi layer. Hence, the resulting structure suggests that the TiSi bilayer acts as a template for the nucleation of \CIVIX, while the formation of C54-like structures would require more drastic intermixing, which is unlikely due to the low diffusivity of Ti interstitials\cite{weber1983} and the rapid increase in defect formation energy beneath the surface. 

Expressed in the C49 lattice basis, the C49(010) plane lies parallel to the Si(100) plane, with the lattice vectors $\mathbf{a}_\mathrm{C49}$ and $\mathbf{c}_\mathrm{C49}$ rotated by \ang{45} with respect to the silicon lattice. The resulting lattice mismatch is \qty{4.18}{\percent}. The diagonal of the Si $\mathbf{a}_\mathrm{Si}\mathbf{b}_\mathrm{Si}$-plane is \qty{7.73}{\Ang}, slightly larger than twice the C49 lattice constants $a$ and $c$ (each \qty{3.71}{\Ang} at the PBE level). Hence, the TiSi bilayer combines a distinctive structural motif with a favorable crystallographic alignment, supporting its role as a nucleation seed for \CIVIX\ growth.

In the broader context of \CIVIX growth, defect formation energies (\Eform) are critical, as defects and atomic diffusion have been identified as key factors favoring \CIVIX over \CVIV. Experimental studies consistently show that Si diffuses more rapidly than Ti, thereby promoting crystal growth.\cite{raaijmakers1990, jeon1992, shimozaki1997, herner1997, roy2014} During annealing, the \TiSix layer thus extends deeper into the silicon bulk, while excess Si forms a surface layer.\cite{taubenblatt_silicide_1982} To better understand this process, Brown \emph{et al.}\,\cite{brown2021} computed bulk defect energetics for both polymorphs, reporting that Si and Ti vacancies are roughly \qty{1}{\eV} to \qty{1.5}{\eV} more favorable in C49 than in C54. We have repeated those calculations for bulk C49 within our computational setup and find that the defect formation energies remain positive with $\Eform(V_\mathrm{Si}^\mathrm{bulk})=\qty{1.07}{\eV}$ and  $\Eform(V_\mathrm{Ti}^\mathrm{bulk})=\qty{2.99}{\eV}$, see \cref{tab:C49defects}.

In contrast, the TiSi bilayer exhibits much lower defect formation energies, with Si vacancies even becoming energetically favorable ($\qty{-0.05}{\eV}$ and $\qty{-0.12}{\eV}$; see \cref{tab:C49defects}). Similarly, $\Eform(V_\mathrm{Ti})$ drops to \qty{0.75}{\eV}, significantly below the bulk value of \qty{2.99}{\eV}. We attribute this reduction to strain induced by lattice mismatch and surface effects, which vacancies help to relieve. The TiSi bilayer thus provides an accessible pathway for atom migration, consistent with experimental observations.

\begin{table}
    \centering
    \begin{tabular}{l r r r r}
    \hline
         defect &  \multicolumn{2}{c}{layer} & \multicolumn{2}{c}{\CIVIX} \\
         & \multicolumn{1}{c}{top} & \multicolumn{1}{c}{subsurface} & \multicolumn{1}{c}{A/C} & \multicolumn{1}{c}{B} \\
    \hline
         \ce{Ti_{Si}} & \num{0.77} & \num{-0.68} & \num{0.41} & \num{0.88} \\
         \ce{Si_{Ti}} & \num{1.13} & \num{1.52} & \num{4.49} & -- \\
         \ce{V_{Ti}}  & \num{0.75} & \num{1.65} & \num{2.99} & -- \\
         \ce{V_{Si}}  & \num{-0.05} & \num{-0.12} & \num{1.07} & \num{1.37} \\
         \hline
    \end{tabular}
    \caption{Defect formation energies \Eform\ of titanium antisites (\ce{Ti_{Si}}), silicon antisites (\ce{Si_{Ti}}), titanium vacancies (\ce{V_{Ti}}), and silicon vacancies (\ce{V_{Si}}) in the TiSi bilayer. For comparison, \Eform\ in bulk \CIVIX\ is included. All energies are in \unit{\eV}.}
    \label{tab:C49defects}
\end{table}

\subsection{Discussion}

In the broader context of Ti growth on Si, our results indicate that initial deposition first populates adsorption sites along the dimer ridges: any nearest-neighbor pairing carries an energetic penalty because it forces open a surface dimer. The decisive step for further growth is the accumulation of a fraction of Ti atoms into the first subsurface layer, which locally lifts the surface reconstruction and renders previously unfavorable adsorption sites accessible. Although the minimum energy path for submerging Ti is not explicitly calculated here, the observation of minor intermixing of Ti and Si in the order of 1 to 3 monolayers prior to annealing suggests that \i--\vi pairs eventually form.\cite{murarka1980,van_loenen1985, vahakangas1986} Given that Ti is a slow diffuser in Si,\cite{weber1983} this process is potentially aided by the dynamic flip-flop motion of surface dimers, which undergo a continuous transition between the $p(2\times2)$ and $c(4\times2)$ surfaces already at low temperatures.\cite{shigekawa1996,neergaard1995,yokoyama2000,kondo2000,hata2002,guo2014} Nevertheless, determining the adsorption kinetics remains a question for future research.  

Once the TiSi bilayer is complete, the growth mode changes abruptly. Additional Ti adsorbates bind much more weakly to the TiSi surface ($\Eads=\qty{+1.43}{\eV}$), whereas Ti-Ti interactions outweigh surface binding ($\Eassc(2\times\text{Ti@TiSi})=\qty{-0.47}{\eV}$; see \SuppInf, Section~3). This promotes Ti clustering irrespective of the substrate’s local orientation, consistent with Stranski-Krastanov growth,\cite{stranski1937,baskaran2012,vahakangas1986,palacio2007} in which a uniform wetting layer is followed by island formation. Compared with earlier computational models,\cite{briquet_first_2013} our results indicate a clear-cut switch in growth mode. Importantly, weakly attractive association energies at low Ti coverage do not necessarily signal the onset of clustering; they can also reflect local cooperative effects. The abrupt change in adsorption behavior once the bilayer is complete rather points to a transition driven by the reduced surface affinity of Ti for the TiSi surface.

These findings suggest a practical route to mitigate the undesired formation of \CIVIX. The TiSi bilayer on Si(100) constitutes an epitaxial template: its zigzag Ti–Si chains reproduce the A/C layers of C49, with C49(010)$\parallel$Si(100), $\mathbf{a}_\mathrm{C49}$ and $\mathbf{c}_\mathrm{C49}$ rotated by \ang{45}, and only \qty{4.18}{\percent} lattice mismatch. Accordingly, our model predicts out-of-plane growth along C49[010], consistent with experiment.\cite{byun_epitaxial_1996} Reports of alternative orientations under rapid thermal annealing (up to \qty{800}{\degreeCelsius})\cite{wan_texture_1996,la_via2000,ozcan2002} and the identification of C49(131) as the lowest-energy surface\cite{iannuzzi2001} indicate that at elevated temperatures kinetics (vacancy formation, interfacial strain relaxation, and surface migration) can override this epitaxial preference.%

If Si(100) seeds \CIVIX via this template, then breaking the surface structure before annealing should suppress \CIVIX, because if the ridge-trench morphology is lost, the driving force for the anisotropic embedding of Ti along the [110] direction disappears as well. In the absence of directionality, the \TiSi growth is then governed by thermodynamic phase stability, favoring the formation of \CVIV. Consistent with this mechanism, it has been observed that preamorphization of Si(100) promotes \CVIV\cite{lu1991,zhu1993,karlin1996,kittl1998,tan2002,yu2016} provided it precedes the \TiSix\ formation step.\cite{motakef1991} Inserting thin refractory interlayers (Mo, Ta, Nb, W) yields similar outcomes,\cite{mann1995,mouroux1996,cabral1997,kittl1998b,mouroux1999,aberg2001} either through blocking adsorption sites and disruption the Ti deposition pattern, or through forming a mixed $\ce{MSi_{\mathit{x}}}$ phase\cite{weber1983,slaughter1991, plyushcheva2009} thereby altering the adsorption energetics. Thus, any perturbation to the Si(100) surface would result in suppressing the surface-topology–directed nucleation of \CIVIX.

\section{Conclusions}

In summary, we have comprehensively modeled the adsorption of Ti on \mySurf, from isolated defects in the dilute limit up to coverages corresponding to two monolayers. For isolated Ti, the preferred adsorption site is in the fourfold hollow site between two adjacent Si surface dimers rather than in the trenches between them. The interactions between Ti adsorbates are mostly repulsive, giving rise to a uniform initial growth of \TiSix; yet Ti interstitials in the first subsurface layer form a strong bond with Ti adsorbates on the surface. These Ti–Ti pairs disrupt the Si dimers and locally lift the surface reconstruction. As a result, the trench adsorption sites become symmetrically equivalent to the favorable adsorption sites, thereby creating a self-promoting mechanism in which higher coverage leads to the formation of more favorable sites and Ti–Ti pairs.

The resulting TiSi bilayer reproduces characteristic low-symmetry motifs of the \CIVIX structure that are unique to this particular polymorph. This suggests that the initial formation of \ce{TiSi2} is governed primarily by the topology of the reconstructed Si(100) surface: the arrangement of buckled Si dimers dictates the initial adsorption, while the spacing between the crystallographic sites in Si are just the right size to embed Ti atoms in between. Eventually, the initial Ti adatom pairs in conjunction with the opening of the Si dimers, that guide the adsorption process towards the characteristic bilayer structure. 

The presented model provides a unified explanation for the initial formation of a Ti rich \TiSix layer at low temperatures, the Stranski-Krastanov growth mode, and offers a plausible mechanism by which the Si(100) surface serves as the nucleation template for \CIVIX, despite its thermodynamic disadvantage relative to \CVIV. This interpretation also aligns with experimental findings that disrupting the surface structure through preamorphization or refractory metals inhibits the C49 formation.
However, our model only includes the first two monolayers and therefore cannot establish a fully quantifiable link. Resolving this will require future studies that extend to higher Ti exposure to model the interface of thicker \CIVIX films on Si(100). Nevertheless, this novel understanding of the unique growth mechanism of \TiSi aids the targeted design of smooth metal-semiconductor junctions to pave the way for next-generation 3D-architectures and photocatalysts.

\section{Computational Methods}
\label{sec:Methods}

All DFT calculations were performed spin-polarized using CP2K\cite{kuhne2020,iannuzzi2025} employing the PBE exchange-correlation functional\cite{perdew1996, perdew1997} with the DZVP-MOLOPT basis set\cite{VandeVondele2007} and the GTH pseudopotentials\cite{Goedecker1996, Hartwigsen1998} to describe the core electrons. The plane-wave cut-off and relative cut-off were converged to \qty{700}{\Ry} and \qty{80}{\Ry} ensuring the SCF convergence to an accuracy of \qty{1e-7}{\eV}, including only the $\Gamma$-point for Brillouin zone sampling. Fermi–Dirac smearing with an electronic temperature of \qty{100}{\K} was applied to aid SCF stability during the transition from semiconducting Si to metallic \TiSix. Bader charges were calculated using the code developed by the Henkelman group.\cite{arnaldssonnodate, henkelman2006,sanville2007,tang2009, yu2011}

\begin{figure}
    \centering
    \includegraphics[scale=1]{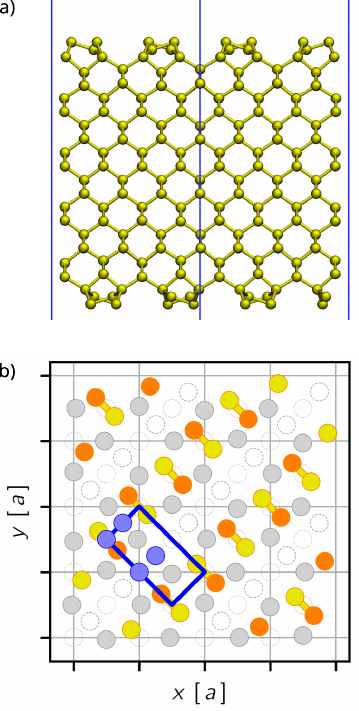}
    \caption{a) The \mySurf slab viewed from the [110]-direction b) Schematic top view of the \mySurf slab with the adsorption sites and the reduced cell used for the exhaustive sampling with \texttt{icet} code.\cite{angqvist2019}}
    \label{fig:icet}
\end{figure}

All Si slab models were prepared using the \texttt{ASE}\cite{hjorth_larsen2017} package based on bulk lattice constants from a fully optimized  3D periodic $3\times3\times3$ cells. The silicon slabs were converged to the surface energy $\Esurf<\qty{0.3}{\eV\per\Ang\squared}$ requiring a thickness of 20 atomic layers (10 crystallographic layers) and lateral expansion of $4\times4$ in the $[100]$- and $[010]$-directions leading to a total of \qty{640}{atoms}. A vacuum spacing of \qty{40}{\Ang} was introduced along the surface normal ($[001]$-direction). The \mySurf surface reconstruction was chosen as the basis for all adsorption configurations as this is the lowest-energy surface configuration according to literature\cite{ramstad1995,guo2014} and our benchmark (see \cref{fig:icet}a, \SuppInf Section~1). Configurations with more than two adatoms were systematically enumerated using the \texttt{icet} code\cite{angqvist2019} with a reduced cell for the adsorbates in accordance with the isolated adsorbates as presented in \textcolor{black}{Section 3.1} %
below. The reduced cell displayed in \cref{fig:icet}b is expanded up to an interaction range of 2, thereby capturing all relevant next-nearest-neighbor interactions as determined in \textcolor{black}{Section 3.2}%
. Structures were visualized using VMD.\cite{humphrey_vmd_1996}

The average adsorption energies per adsorbate (\Eads) are calculated via 
\begin{align}
    \Eads=\frac{E_{n\ce{Ti}@\ce{Si}}-E_\mathrm{slab}-n\mu_{\ce{Ti}}}{n}
\end{align}
where $\mu_{\ce{Ti}}$ is the chemical potential of Ti, referenced to bulk $\alpha$-Ti.  While earlier works\cite{anez2012, briquet_first_2013} used isolated Ti atoms in vacuum for $\mu_{\ce{Ti}}$ to approximate conditions of atomic layer deposition (ALD), we adopt the bulk reference ($\mu_{\ce{Ti}}=E_\mathrm{bulk}(\alpha\text{-Ti})/N_\mathrm{Ti}$) to better assess the competition between Ti–Si binding and Ti–Ti clustering. As a result, our reported \Eads values are shifted by the cohesive energy of bulk $\alpha$-Ti, approximately \qty{-5.27}{\eV}, relative to the quoted studies.
The association energy (\Eassc) is defined by comparing \Eads of a system with $N$ adsorbates to the sum of isolated adsorbates:
\begin{align}
\Eassc = \Eads^{N} - \sum_i^N E_{\mathrm{ads},i}^{\mathrm{isolated}}~,
\end{align}
where a negative value indicates an energetic preference for clustering and a positive value repulsion among adsorbates.

\begin{acknowledgement}
This work was conducted at the Advanced Research Center for Nanolithography, a public-private partnership between the University of Amsterdam (UvA), Vrije Universiteit Amsterdam (VU), Rijksuniversiteit Groningen (RUG), the Netherlands Organization for Scientific Research (NWO), and the semiconductor equipment manufacturer ASML. 
E.O. is grateful for a WISE Fellowship from the NWO. J.M. and E.O. acknowledge support via Holland High Tech through a public–private partnership in research and development within the Dutch top sector of High-Tech Systems and Materials (HTSM). The use of the national computer facilities in this research was subsidized by NWO Domain Science (grant no. 2024.029). %
 The authors thank SURF (www.surf.nl) for the support in using the National Supercomputer Snellius.

\end{acknowledgement}

\begin{suppinfo}
Additional Computation Details for Bulk \cSi, Si(100) Surface and Bulk \CIVIX,
Ti Interstitial Defects Close to the Si(100) Surface, Ti Adsorption on the TiSi Bilayer.

All structures generated in this study are available via \texttt{zenodo.org} at \url{https://doi.org/10.5281/zenodo.18242845}.
\end{suppinfo}
\bibliography{bibliography}

\end{document}